\documentclass{article}
\usepackage{graphicx} 
\usepackage{amsmath,amssymb,amsthm}
\usepackage{parskip}
\usepackage{natbib}
\usepackage{float}
\usepackage[colorlinks]{hyperref}
\usepackage[margin=2cm]{geometry}
\usepackage{tikz, tikz-cd}
\usepackage[color=yellow]{todonotes}
\usepackage{array}
\usepackage{authblk}

\newtheorem{theorem}{Theorem}

\def\contract{\makebox[1.2em][c]{\mbox{\rule{.6em}
{.01truein}\rule{.01truein}{.6em}}}}

\title{Symmetry groups of geodesic equations with applications in water waves}
\author[1]{Bo Gervang}
\author[2]{Erwin Luesink}
\affil[1]{Department of Engineering Technology, Technical University of Denmark, Lautrupvang 15, DK-2750 Ballerup, Denmark}
\affil[2]{Korteweg-De Vries Institute, University of Amsterdam, PO Box 94248, Science Park 107, 1090 GE Amsterdam, The Netherlands}
\date{August 2024}

\begin{document}

\maketitle
\begin{abstract}
    In this work we derive several important equations in water waves and liquid crystals by deriving them as geodesic equations of right-invariant metrics on two infinite-dimensional groups. The equations we obtain this way are the Hopf (inviscid Burgers) equation, the Camassa-Holm equation, the Hunter-Saxton equation and the Korteweg-De Vries equation. We then study the symmetry groups of the equations themselves and show that one can improve the behaviour of the Hopf equation by metric and topological corrections. The symmetry groups of these equations can aid the benchmarking and testing of numerical methods.
\end{abstract}

\section{Introduction}
After the seminal work \cite{arnold1966geometrie}, ideal fluid dynamics gained an interpretation in geometric terms: the Euler equations of ideal incompressible fluid dynamics are the geodesics of the volume-preserving diffeomorphism group over a compact manifold in the $L^2$-norm. While this result is independent of the dimension of the underlying compact manifold as long as the dimension is larger than or equal to two, the one-dimensional case is special. Indeed, in one dimension, the only incompressible flows are the constant flows. Hence, for the one-dimensional case, it is natural to study compressible flows, which means that the group in this case is the entire diffeomorphism group. The only compact manifold in one dimension is the circle $S^1$, so the study of one-dimensional fluid dynamics in the sense of \cite{arnold1966geometrie} concerns the unidirectional diffeomorphism group $\mathfrak{D}^+$, which is the connected component of the identity of the diffeomorphism group, over the circle. 

In dimension one, there are two smooth Riemannian manifolds: the noncompact real line, which is homeomorphic to $\mathbb{R}$ in a trivial manner, and the circle $S^1 = \mathbb{R}/\mathbb{Z}$, which is locally homeomorphic to $\mathbb{R}$. Though the circle is the lowest dimensional compact manifold one can study, the diffeomorphism group over the circle has a rich mathematical nature that leads to several remarkable partial differential equations. A key difference with higher dimensional manifolds is that in dimension one every 1-form is exact, which means that fluid dynamics in dimension one only considers potential flows. This is fundamentally different from the higher dimensional case, where one can use the Hodge decomposition to dissect fluid dynamics into three components: the divergence-free flows, which are associated with vorticity, the irrotational flows, which are associated with waves, and the harmonic flows, which are associated with the kernel of the Hodge Laplacian. In dimension one, the Hodge decomposition shows that a divergence-free flow is constant, meaning that the vorticity is always zero. The irrotational component is not trivial so dimension one is suitable for studying waves.

The geodesic equation for the $L^2$-norm on the diffeomorphism group over the circle is the Hopf equation (or the inviscid Burgers equation). This equation is known to form shocks. The shock is a transition from a strong solution to a weak solution in our setting. There are several ways to change the solution behaviour and we explore two such ways. One well-known regularisation that we do not use is the addition of viscous dissipation. This regularisation would lead to the Burgers equation, which dissipates energy and can not be formulated as a geodesic equation. The first way of regularising the Hopf equation that we explore is a particular change of metric, and we show that this leads to the Camassa-Holm equation and the Hunter-Saxton equation. The second way is the improvement of the topology of the diffeomorphism group over the circle. This improvement uses a universal central extension and changes the group. The universal central extension of the diffeomorphism group over the circle is the Virasoro-Bott group and its $L^2$-geodesics are described by the Korteweg-De Vries equation. 

The Korteweg-De Vries equation was named after the paper \cite{korteweg1895xli}, the Camassa-Holm equation after the paper \cite{camassa1993integrable} and the Hunter-Saxton equation after the paper \cite{hunter1994completely}. Besides the study of the symmetry groups involved in the derivation of the equations, and the study of the symmetry groups of the equations themselves, this work also celebrates the upcoming 130th birthday of the Korteweg-De Vries equation, as well as the recent 30th birthdays of the Camassa-Holm and Hunter-Saxton equations.

Table \ref{tab:introtable}, parts of which can also be found in \cite{khesin2008geometry}, gives an overview of some of the geodesic equations that are obtained by equipping the groups $\mathfrak{D}^+$ and $\mathrm{Vir}$ with metrics $L^2$, $H^1$ and $\dot{H}^1$ (the $H^1$-norm without the $L^2$-part). 

\begin{table}[H]
    \centering
    \begin{tabular}{c|c|c}
        Group & Metric & Equation \\ \hline
        $\mathfrak{D}^+$ & $L^2$ & Hopf\\
        $\mathfrak{D}^+$ & $H^1$ & Camassa-Holm \\
        $\mathfrak{D}^+$ & $\dot{H}^1$ & Hunter-Saxton \\
        $\mathrm{Vir}$ & $L^2$ & Korteweg-De Vries \\
        $\mathrm{Vir}$ & $H^1$ & Dispersive Camassa-Holm \\
        $\mathrm{Vir}$ & $\dot{H}^1$ & Dispersive Hunter-Saxton 
    \end{tabular}
    \caption{Overview of geodesic equations}
    \label{tab:introtable}
\end{table}

In Section \ref{sec:topgemprelim} we introduce the topological and geometric preliminaries to derive the equations listed in Table \ref{tab:introtable}. In Section \ref{sec:diff}, we discuss the group of diffeomorphisms on the circle, in particular its adjoint and coadjoint representations. In Section \ref{sec:ep}, we use the Euler-Poincar\'e theorem to derive geodesic equations for the diffeomorphism group. This leads to the Hopf, Camassa-Holm and Hunter-Saxton equations. In Section \ref{sec:vir} we discuss the universal central extension of the group of diffeomorphisms on the circle, the Virasoro-Bott group. By means of the adjoint and coadjoint representations and the Euler-Poincar\'e theorem, we derive the Korteweg-De Vries equation, the dispersive version of the Camassa-Holm equation and the dispersive version of the Hunter-Saxton equation. In Section \ref{sec:symmetry} we perform symmetry analysis of each of the mentioned equations to analyse how solutions may be transformed to other solutions. Finally, in Section \ref{sec:conclusion} we conclude.

\section{Topological and geometric preliminaries}\label{sec:topgemprelim}
In this section we discuss the topological and geometric tools that are necessary for the derivation of the equations listed in Table \ref{tab:introtable}. These equations are all examples of Euler-Poincar\'e and Lie-Poisson equations (sometimes also called Euler-Arnold equations). Both of these equations are studied in the field of geometric mechanics. The Euler-Poincar\'e equations correspond to Euler-Lagrange equations after symmetry reduction and the Lie-Poisson equations correspond to the canonical Hamilton's equations after symmetry reduction. These relations are best captured in the diagram below.
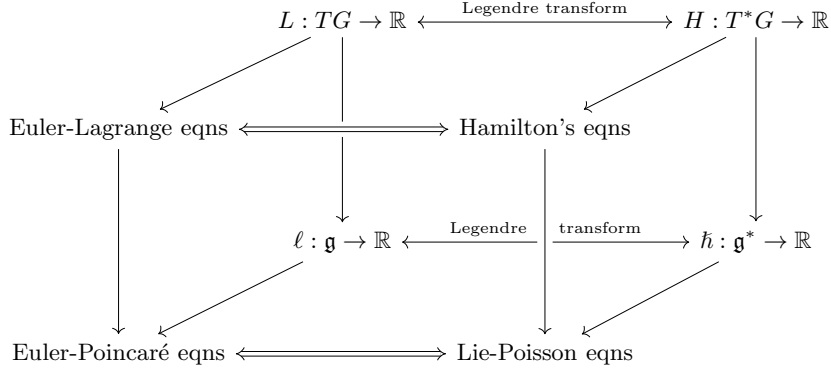
\begin{figure}[H]
\small
\centering
\begin{tikzcd}[row sep=3em, column sep=small]
& 
L:TG\to\mathbb{R} \arrow[dl] \arrow[rr,  "\text{Legendre transform}", leftrightarrow] \arrow[dd] 
& 
& 
H:T^*G\to\mathbb{R} \arrow[dl] \arrow[dd]
\\
\text{Euler-Lagrange eqns} \arrow[rr, crossing over, Leftrightarrow] 
& 
& \text{Hamilton's eqns}
\\
& \ell:\mathfrak{g}\to\mathbb{R} \arrow[dl] \arrow[rr, "\text{Legendre \hspace{0.25cm} transform}", leftrightarrow] 
& 
& \hslash:\mathfrak{g}^*\to\mathbb{R} \arrow[dl]
\\
\text{Euler-Poincar\'e eqns} \arrow[rr, Leftrightarrow] \arrow[from=uu, crossing over]
& 
& \text{Lie-Poisson eqns} \arrow[from=uu, crossing over]
\end{tikzcd}
\caption{The cube of commutative diagrams for geometric mechanics on Lie groups. Euler-Poincar\'e reduction (on the left side) and Lie-Poisson reduction (on the right side) are both indicated by the arrows pointing down. The diagrams are all commutative, provided the Legendre transformations are invertible.}
\label{fig:cube}
\end{figure}
In the diagram, starting with the top left corner in the back, $G$ denotes a Lie group or a topological group and $TG$ is the tangent bundle over $G$. In general, $G$ is allowed to be a smooth manifold, but we focus only on the case where $G$ is a Lie group or a topological group. The Lagrangian $L:TG\to\mathbb{R}$ is a functional that is usually given by kinetic energy minus potential energy. In the top right corner in the back, $T^*G$ is the cotangent bundle and the Hamiltonian $H:T^*G\to \mathbb{R}$ is a functional that is usually the sum of kinetic energy and potential energy. By means of the Legendre transform, which is a diffeomorphism if the Lagrangian and/or Hamiltonian is hyperregular (this means that the functionals have nonconstant fiber derivatives), one can transform between Lagrangian descriptions and Hamiltonian descriptions. On the Lagrangian side, via a variational principle known (slightly confusingly) as Hamilton's principle, one obtains the Euler-Lagrange equations. On the Hamiltonian side, via a Poisson bracket or a symplectic structure, one obtains Hamilton's canonical equations. 

Groups are equipped with a group action, which can be tangent lifted to give a natural action on the tangent bundle of the group and similarly, the group action can be cotangent lifted to give a natural action on the cotangent bundle of the group. If the Lagrangian or Hamiltonian is invariant under the appropriate (left or right) action, then symmetry reduction is possible. For the Hamiltonian version, the symmetry reduction method was introduced by \cite{marsden1974reduction} and is called Lie-Poisson reduction. On the Lagrangian side, symmetry reduction was introduced quite a bit later by \cite{holm1998euler} and is called Euler-Poincar\'e reduction. Euler-Poincar\'e reduction takes the Lagrangian functional on the tangent bundle and associates a reduced Lagrangian functional $\ell:\mathfrak{g}\to\mathbb{R}$ on Lie algebra $\mathfrak{g}$ associated with the underlying group. Similarly, Lie-Poisson reduction takes the Hamiltonian functional on the cotangent bundle and associates a reduced Hamiltonian functional $\hslash:\mathfrak{g}^*\to\mathbb{R}$ on the dual $\mathfrak{g}^*$ of the Lie algebra. 

To go from the reduced Lagrangian to equations of motion, Hamilton's variational principle needs to be adapted to account for the symmetries. The adapted variational principle is called the Euler-Poincar\'e variational principle and upon applying this to the Lagrangian, one obtains the Euler-Poincar\'e equations. On the Hamiltonian side, one needs to adapt the Poisson bracket or symplectic structure. The appropriate bracket is called the Lie-Poisson bracket and together with the reduced Hamiltonian yields the Lie-Poisson equations. This completes a general explanation of the cube in Figure \ref{fig:cube}. 

For a detailed discussion of finite-dimensional geometric mechanics, a number of textbooks are available, for instance \cite{abraham1978foundations, holm2008geometric1, holm2008geometric2, holm2009geometric, marsden2013introduction}. For our purposes, we do not develop the general theory in full detail, but shall discuss only the diagram when the group is the group of diffeomorphisms on the circle. In infinite dimensions, the different topologies are not equivalent, which leads to numerous technicalities and subtleties. Most importantly, Lie groups in the infinite-dimensional setting are modelled on Fr\'echet spaces (which are complete locally convex Hausdorff metrisable vector spaces). Such Lie groups are called Fr\'echet Lie groups. While defining tangent spaces, vector fields, differential forms and such is for all intents and purposes the same as in the finite dimensional case, however, there are several important differences. The dual of a Fr\'echet space need not be Fr\'echet (in fact the dual of a Fr\'echet space is itself a Fr\'echet space if and only if it is a Banach space, see \cite{kriegl1997convenient}), which in particular means that differential forms have to be defined directly. Let $\mathfrak{g}$ be the Lie algebra of a Fr\'echet Lie group $G$, then the Lie algebra is a Fr\'echet space. In case the dual is not again a Fr\'echet space, one can confine themselves to considering appropriate smooth duals, which are functionals from a certain group invariant Fr\'echet subspace $\mathfrak{g}^*_s\subset\mathfrak{g}^*$. 

The diffeomorphisms over the circle are a Fr\'echet Lie group. One can also consider manifolds and Lie groups modelled on Banach spaces, leading to Banach manifolds and Banach Lie groups. These have the advantage that powerful results from finite-dimensional analysis hold, such as a general solution theory for differential equations, and the notion of duality is the usual one. However, diffeomorphism groups are not Banach manifolds. It is possible to enlarge the class of diffeomorphisms by considering the $C^k$-topology or the $H^s$ topology, which lead to Banach manifolds and Hilbert manifolds, respectively, but such manifolds are no longer Lie groups as a result of the following. While the right action is smooth, the left action is only continuous (not even Lipschitz continuous) by theorems of \cite{palais1966foundations} and \cite{ebin1967space}. Although the groups are merely topological groups in this case, the cube in Figure \ref{fig:cube} remains valid as long as one only considers right actions and right-invariant functionals. Fortunately, this is precisely the setting for the type of geodesic equations of interest in this work, meaning that we can use the Fr\'echet, Banach and Hilbert frameworks for the analysis. In the next section, we introduce the necessary theory to discuss geodesics on the group of diffeomorphisms on the circle.

\section{Diffeomorphisms on the circle}\label{sec:diff}
For a detailed discussion of the group of diffeomorphisms on the circle we refer to \cite{khesin2008geometry}. Here, we recall the basic operations that are relevant for the derivation of geodesic equations on this group.

By diffeomorphism group on the circle $\mathrm{Diff}(S^1)$ we mean the Fr\'echet Lie group of smooth functions whose domain and image are the circle and whose inverse is also smooth. The Lie algebra $\mathfrak{X}(S^1)$ of the diffeomorphism group of the circle consists of the smooth vector fields on the circle. Both the group and the algebra admit central extensions, which improve the coadjoint representation compared to the nonextended versions. We now explain what coadjoint orbits are for the group of diffeomorphisms, after which we explain the notion of a central extension and how it improves coadjoint representations.

We first introduce the diffeomorphism group in a more precise manner. If $M$ and $N$ are two compact differentiable manifolds, a diffeomorphism $f:M\to N$ is a smooth bijection whose inverse $f^{-1}:N\to M$ is also smooth. The set of all diffeomorphisms from $M$ to $N$ is denoted $\mathrm{Diff}(M,N)$. When $M=N$, we can equip $\mathrm{Diff}(M,M)=\mathrm{Diff}(M)$ with a group action. The natural choice of a group product is composition $\circ:\mathrm{Diff}(M)\times\mathrm{Diff}(M)\to \mathrm{Diff}(M)$ and this leads to the group of diffeomorphisms over $M$. The lowest-dimensional setting for diffeomorphism group over a compact differentiable manifold happens in one dimension when $M=S^1$. This group is denoted $\mathrm{Diff}(S^1)$ and consists of two connected components that we denote by $\mathfrak{D}^+(S^1)$, the subgroup of orientation-preserving diffeomorphisms over the circle, and $\mathfrak{D}^-(S^1)$, the space of orientation-reversing diffeomorphisms. Since the identity preserves orientation, only $\mathfrak{D}^+(S^1)$ is a subgroup, because $\mathfrak{D}^-(S^1)$ lacks an identity element. The orientation-preserving diffeomorphisms of the circle $\mathfrak{D}^+(S^1)$ and its Lie algebra of vector fields on the circle $\mathfrak{X}(S^1)$ are the subject of study in this work.

For the diffeomorphism group the exponential map $\mathfrak{X}(M)\mapsto \mathrm{Diff}(M)$ has to assign to each vector field on $M$ the map for its flow. Such a map may not exist when $M$ is not compact. Since the circle is compact, the exponential map is well defined. However, it can be shown that the exponential map $\exp:\mathfrak{X}(S^1)\to \mathrm{Diff}(S^1)$ is not locally surjective. This can be shown by means of the following argument. Observe that any nowhere-vanishing vector field on $S^1$ can be transformed to a constant vector field by means of a transformation in $\mathrm{Diff}(S^1)$. Let $\xi = f(x)\partial_x$ be such a vector field. We define the diffeomorphsim $\psi:S^1\to S^1$ as $\psi(x) = c \int_0^x \frac{ds}{f(s)}$, where $c\in \mathbb{R}$ is a constant such that $\psi(2\pi)=2\pi$. Then the push-forward $\psi_*(\xi\circ\psi^{-1})$ is a constant vector field on the circle. This means that any diffeomorphism of $S^1$ that lies in the image of the exponential map without fixed points is conjugate to a rigid rotation of the circle. To show that the exponential map is not locally surjective, it suffices to construct diffeomorphisms arbitrarily close to the identity that do not have any fixed points and that are not conjugate to a rigid rotation. This can be achieved by constructing diffeomorphisms with isolated periodic points (fixed points for some $n$th iteration of this diffeomorphism). 

To understand the inner workings of the group $\mathfrak{D}^+(S^1)$, we study its representation theory. In what follows, the underlying manifold is always the circle, so for notational convenience we write $\mathfrak{D}^+$ and $\mathfrak{X}$ for the diffeomorphism group over the circle and the space of vector fields over the circle, respectively. In general, a representation of a Lie group on a vector space $V$ is a linear action $\varphi$ of the group $G$ on $V$ for which it holds that the map $G\times V\to V$ given by $(g,v)\mapsto gv$ is smooth. Every Lie group has two particularly important representations: the adjoint representation and the coadjoint representation. We first introduce the concepts in an abstract manner for the diffeomorphisms over the circle, after which we compute them explicitly in coordinates.

Any element in $\psi \in \mathfrak{D}^+$ defines an automorphism $\mathrm{AD}_\psi$ of the group by conjugation
\begin{equation}\label{def:automorphism}
    \mathrm{AD}_{\psi}(\phi) := \psi\circ\phi\circ \psi^{-1},
\end{equation} 
where $\circ$ denotes composition. The group of automorphisms of the diffeomorphism group over the circle is denoted by $\mathrm{Aut}(\mathfrak{D}^+)$. The automorphism in particular is a map $\mathrm{AD}:\mathfrak{D}^+\times \mathfrak{D}^+ \to \mathfrak{D}^+$. By differentiating $\mathrm{AD}_\psi$ at the identity $e\in\mathfrak{D}^+$ one obtains a map from $\mathfrak{X}$ to itself
\begin{equation}
    \mathrm{Ad}_{\psi}(X) :=  \frac{d}{dt}\Big|_{t=0}\mathrm{AD}_{\psi}(\exp tX).
\end{equation}
Note that $\mathrm{Ad}_\psi:\mathfrak{X}\to\mathfrak{X}$ and thus lives in $\mathrm{Aut}(\mathfrak{X})$, the automorphism group of the Lie algebra. It can be shown that $\mathrm{Ad}$ is a group representation. The orbits of the group in its Lie algebra given by $\{\mathrm{Ad}_\psi X \,\vert\, X\in\mathfrak{X}, \psi\in\mathfrak{D}^+ \}$ are called the adjoint orbits. The differential of $\mathrm{Ad}:\mathfrak{D}\to\mathrm{Aut}(\mathfrak{X})$ at the identity defines a map $\mathrm{ad}:\mathfrak{X}\to\mathrm{End}(\mathfrak{X})$, where $\mathrm{End}(\mathfrak{X})$ denotes the space of endomorphisms of $\mathfrak{X}$. The map $\mathrm{ad}$ is called the adjoint representation of the Lie algebra $\mathfrak{X}$, i.e.,
\begin{equation}
    \mathrm{ad}_X(Y) := \frac{d}{dt}\Big|_{t=0}\mathrm{Ad}_{\exp (tX)}(Y),
\end{equation}
for $X,Y\in\mathfrak{X}$. The Lie algebra $\mathfrak{X}$ of vector fields is a module (i.e., a vector space over a ring), for which the dual module is defined as the space of homomorphisms $\mathfrak{X}^* := \mathrm{Hom}(\mathfrak{X}^*,\mathbb{R})$. Since the dual $\mathfrak{X}^*$ of a Fr\'echet space $\mathfrak{X}$ is not necessarily again a Fr\'echet space, one may have to restrict to the subspace of smooth duals $\mathfrak{X}^*_s$. The coadjoint representation $\mathrm{Ad}^*_s=\mathrm{Ad}^*|_{\mathfrak{g}^*_s}$ of $\mathfrak{D}^+$ of the group on the space $\mathfrak{X}_s^*\subset \mathfrak{X}^*$ can now be defined. We abuse notation and neglect the subscript $s$ indicating this subset. Let $\langle \, ,\,\rangle$ denote the pairing between $\mathfrak{X}$ and its dual $\mathfrak{X}^*$, then the coadjoint action $\mathrm{Ad}^*_\psi:\mathfrak{X}^*\to\mathfrak{X}^*$ is defined by
\begin{equation}\label{def:coadjointgroup}
    \langle \mathrm{Ad}^*_\psi(\mu), X\rangle := \langle \mu, \mathrm{Ad}_{\psi^{-1}}(X)\rangle
\end{equation}
for all $\mu\in\mathfrak{X}^*$ and $X\in\mathfrak{X}$. Note that the definition of the coadjoint action includes the inverse of the group element. This guarantees that $\mathrm{Ad}^*$ is also a group representation. The differential $\mathrm{ad}^*:\mathfrak{X}\to\mathrm{End}(\mathfrak{X}^*)$ of the group representation $\mathrm{Ad}^*:\mathfrak{D}^+\to \mathrm{Aut}(\mathfrak{X}^*)$ at the group identity, i.e.,
\begin{equation}
    \mathrm{ad}^*_X \mu := \frac{d}{dt}\Big|_{t=0} \mathrm{Ad}^*_{\exp(tX)} \mu,
\end{equation}
for $\mu\in\mathfrak{X}^*$ and $X\in\mathfrak{X}$ is called the coadjoint representation of the Lie algebra. Equivalently, the coadjoint representation of the Lie algebra can be obtained via duality from the adjoint representation of the Lie algebra by the relation
\begin{equation}
    \langle \mathrm{ad}^*_Y(\mu), X\rangle := - \langle \mu, \mathrm{ad}_Y(X)\rangle.
\end{equation}
This concludes the discussion of adjoint and coadjoint representations in their abstract form. Next, we compute these operators explicitly. The automorphism is given explicitly in its definition \eqref{def:automorphism} by conjugation. For the vector fields on the circle we can give an explicit basis. Fixing an angle $x$, any vector field $X\in\mathfrak{X}$ can be written as $X(x) = u(x)\partial_x$ with $u(x)\in C^\infty(S^1)$. By differentiation and using the chain rule, the adjoint representation of the group is computed to be
\begin{equation}
\begin{aligned}
    \mathrm{Ad}_{\psi}(X(x)) &= \frac{d}{dt}\Big|_{t=0} \mathrm{AD}_\psi (\exp t X(x))\\
    &= \frac{d}{dt}\Big|_{t=0} \psi \circ \exp(t X(\psi^{-1}(x))\\
    &= \psi \circ X(x) \circ \psi^{-1}\\
    &= \psi_* X(x) \\
    &= X(\psi).
\end{aligned}
\end{equation}
The adjoint representation of the Lie algebra can be computed by differentiating once more
\begin{equation}
    \begin{aligned}
        \mathrm{ad}_X(Y) &= \frac{d}{dt}\Big|_{t=0} \mathrm{Ad}_{\exp tX}(Y) \\
        &= \frac{d}{dt}\Big|_{t=0} \exp(tX)_* Y\\
        &= -\mathcal{L}_X Y \\
        &= [X,Y],
    \end{aligned}
\end{equation}
where $\mathcal{L}_X Y$ denotes the Lie derivative of $Y$ along $X$ and $[X,Y]$ is the Jacobi-Lie bracket of vector fields. In the coordinate expression with $X(x)=u(x)\partial_x$ and $Y(x) = v(x)\partial_x$, the adjoint representation of the Lie algebra takes the form
\begin{equation}
    \mathrm{ad}_{u(x)\partial_x}(v(x)\partial_x) = (uv_x - u_x v)\partial_x,
\end{equation}
where we have suppressed the arguments of $u$ and $v$ on the right-hand side and $u_x$ denotes $\partial u/\partial x$. Next we compute the duals of both the adjoint representations to obtain the coadjoint representations. For this we require an explicit understanding of the (smooth) dual space of the vector fields on the circle. Here it helps to view the vector fields as smooth sections of the space of vector-valued scalars, i.e., as a special case of bundle-valued differential forms, elements of $\mathfrak{X}=\Omega^0(S^1,TS^1)$ . The smooth dual space of the vector fields is the space of covector-valued densities, which are elements of the space $\mathfrak{X}^*= \Omega^1(S^1, T^*S^1)$. Note that there is two levels of duality here. Firstly, fixing the base point in $S^1$, a vector in each fiber is dual to a covector. Secondly, a scalar function on $S^1$ is dual to a density on $S^1$. Putting these two notions of duality together lets us identify the smooth dual space of vector fields with the space of covector-valued densities. In the explicit coordinate form, covector-valued densities $\mu\in \mathfrak{X}^*$ have the form $\mu(x) = m(x)dx\otimes dx$ with $m(x)\in C^\infty(S^1)$. The pairing $\langle\,,\,\rangle$ has the explicit form
\begin{equation}
\begin{aligned}
    \langle \mu, X\rangle &= \int_{S^1} X\contract \mu \\
    &= \int_{S^1} u(x)\partial_x\contract m(x)dx\otimes dx \\
    &= \int_{S^1} u(x)m(x) dx.
\end{aligned}
\end{equation}
Here $X\contract \mu$ denotes the interior product between the vector field $X$ and the covector-valued density $\mu$. The adjoint representation of the group of diffeomorphisms on the circle is the push-forward on vector fields, which means that the coadjoint representation of the group is the pull-back of covector-valued densities. We compute the coadjoint action from the definition in \eqref{def:coadjointgroup}, but with the inverses reversed for the sake of interpretation
\begin{equation}
    \begin{aligned}
        \langle \mathrm{Ad}^*_{\psi^{-1}} \mu, X\rangle &= \langle \mu, \mathrm{Ad}_{\psi}X\rangle \\
        &= \langle \mu, \psi_* X\rangle \\
        &= \langle \psi^*\mu, X\rangle.
    \end{aligned}
\end{equation}
In the coordinate expression $\mu(x) = m(x)dx\otimes dx$, we then have
\begin{equation}
    \mathrm{Ad}^*_{\psi^{-1}}(m(x)dx\otimes dx) = m(\psi)d\psi\otimes d\psi.
\end{equation}
Finally, the computation of the coadjoint representation of the Lie algebra can be performed in two ways. We can compute the dual of the adjoint representation of the Lie algebra or differentiate the coadjoint representation of the group. In either case, we obtain the same expression. The computation via duality is as follows
\begin{equation}
\begin{aligned}
    \langle \mathrm{ad}^*_Y \mu, X\rangle &= \langle \mu, -\mathrm{ad}_Y X\rangle \\
    &= \langle \mu, \mathcal{L}_Y X\rangle \\
    &= \langle \mathcal{L}_Y \mu, X\rangle
\end{aligned}
\end{equation}
In the coordinate expression $\mu(x)dx\otimes dx$ and $X(x) = f(x)\partial_x$, we have
\begin{equation}
    \mathrm{ad}^*_{u(x)\partial_x}(m(x)dx\otimes dx) = ((mu)_x + mu_x)dx\otimes dx.
\end{equation}
This is the coordinate expression of the Lie derivative of a covector-valued density in dimension one. In summary, the automorphism, adjoint and coadjoint representations of $\mathfrak{D}^+$ and its Lie algebra $\mathfrak{X}$ are in Table \ref{tab:diffcircle}.
\begin{table}[H]
\centering
\setlength{\extrarowheight}{6pt}
\begin{tabular}{ll}
$\mathrm{AD}_{\varphi}:$ & $ \widetilde{\varphi} \mapsto \varphi\circ\widetilde{\varphi}\circ\varphi^{-1}$ \\
$\mathrm{Ad}_{\varphi}:$ & $ u(x)\partial_x\mapsto u(\varphi)\partial_\varphi$ \\
$\mathrm{Ad}^*_{\varphi^{-1}}:$ & $ m(x)dx\otimes dx\mapsto m(\varphi)d\varphi\otimes d\varphi$ \\
$\mathrm{ad}_{u\partial_x}:$ & $v\partial_x\mapsto (uv_x-vu_x)\partial_x$\\
$\mathrm{ad}^*_{u\partial_x}:$ & $m(x)dx\otimes dx \mapsto ((mu)_x +mu_x)dx\otimes dx$
\end{tabular}
\caption{The automorphism, adjoint and coadjoint actions of the diffeomorphism group over the circle and the vector fields over the circle.}
\label{tab:diffcircle}
\end{table}
Next, we show that the Euler-Poincar\'e theorem, instrumental in Figure \ref{fig:cube}, provides the pathway from a minimisation problem on a Lie group towards differential equations on the dual of a Lie algebra by means of the coadjoint representation on the Lie algebra.

\section{Euler-Poincar\'e equations on diffeomorphism groups}\label{sec:ep}
The Euler-Poincar\'e equations on groups of diffeomorphisms over compact smooth manifolds are encountered in various areas of mathematics and physics. Because they are so common and their name is quite unwieldy in the present form, they are often called EPDiff equations. One of the most important examples of EPDiff equations is the Euler equations of ideal fluid dynamics on smooth manifolds, following the seminal paper \cite{arnold1966geometrie}. In this geometric perspective, the starting point is a (weak) Riemannian metric on the group of diffeomorphisms. A Riemannian metric is a positive definite inner product $g_p$ on each tangent space $T_p M$ of an $n$-dimensional smooth manifold $M$. We consider only smooth Riemannian metrics: for each smooth coordinate chart $(U,x)$ on $M$ the $n^2$ functions given by
\begin{equation}
    g(\partial_{x^i},\partial_{x^j}):U\to\mathbb{R}
\end{equation}
are smooth. A weak Riemannian metric can be obtained in a natural manner on the group of diffeomorphisms over a Riemannian manifold $M$ by integrating the metric against a volume form over the manifold. This metric is weak, since the smoothness of the topology on the base manifold does not matter, one inherits merely the $L^2$-type topology on the group of diffeomorphisms. Alternatively, one can also define metrics directly on groups of diffeomorphisms. For example the $L^2$-metric on $\mathfrak{D}^+$ can be expressed in coordinates as
\begin{equation}\label{eq:l2metric}
    g(\dot{\psi},\dot{\psi}) = \frac{1}{2}\int_{S^1} |\dot{\psi}(x)|^2 dx,
\end{equation}
where $\dot{\psi}$ denotes the element in the fiber $T_\psi\mathfrak{D}^+$. This metric can be viewed as a Lagrangian functional $L(\psi,\dot{\psi}) = g(\dot{\psi},\dot{\psi})$ on the tangent bundle $T\mathfrak{D}^+$. Note that in this case the Lagrangian does not explicitly depend on $\psi$. In this case it can be shown that the Lagrangian (and thereby the metric) is right-invariant under the tangent-lift of the action of $\mathfrak{D}^+$ on $T\mathfrak{D}^+$. This means that we right-trivialise the Lagrangian to obtain a reduced Lagrangian $L(\psi,\dot{\psi})\circ\psi^{-1} = L(e,\dot{\psi}\circ\psi^{-1}) =: \ell(X)$. The reduced Lagrangian $\ell:\mathfrak{X}\to\mathbb{R}$ is a functional on the Lie algebra $\mathfrak{X}$. The Euler-Poincar\'e theorem, given in general form in \cite{holm1998euler}, connects the Euler-Lagrange equations obtained from the original Lagrangian with the Euler-Poincar\'e equations obtained from the reduced Lagrangian. It makes use of the notion of a functional derivative, which is defined through the Gateaux derivative in the following manner. Let $F:\mathcal{B}\to\mathbb{R}$ be any differentiable functional on a Banach space $\mathcal{B}$ (Fr\'echet spaces are possible provided one takes good care of the dual). The functional derivative $\delta F/\delta \psi\in\mathcal{B}^*$ is defined as 
\begin{equation}
\delta F[\psi] = \frac{d}{d\epsilon}\Big|_{\epsilon=0} F[\psi + \epsilon\nu] =: \langle \frac{\delta F}{\delta \psi}, \nu\rangle,
\end{equation} 
where $\nu\in \mathcal{B}$ is arbitrary and $\langle\,,\,\rangle$ denotes the pairing between $\mathcal{B}$ and its dual. The statement of the theorem for right-invariance and right-representations (right/right) on $\mathfrak{D}^+$ is as follows.

\begin{theorem}[Euler-Poincar\'e]
    The following are equivalent:
    \begin{enumerate}
        \item Hamilton's variational principle of least action
            \begin{equation}
                \delta \int_{t_1}^{t_2} L(\psi,\dot{\psi})\,dt = 0
            \end{equation}
            holds for variations $\delta\psi$ of $\psi$ vanishing at the endpoints $t_1$ and $t_2$.
        \item $\psi$ satisfies the Euler-Lagrange equations for $L$ on $\mathfrak{D}^+$.
        \item The constrained variational principle
            \begin{equation}
                \delta\int_{t_1}^{t_2} \ell(X)\,dt = 0
            \end{equation}
            holds on $\mathfrak{X}$, using variations of the form
            \begin{equation}
                \delta X = \dot{Y} - \mathrm{ad}_X Y
            \end{equation}
            where $Y\in\mathfrak{X}$ vanishes at the endpoints.
            \item The Euler-Poincar\'e equation holds on $\mathfrak{X}^*$
                \begin{equation}\label{eq:eulerpoincare}
                    \frac{d}{dt}\frac{\delta\ell}{\delta X} + \mathrm{ad}^*_{X}\frac{\delta\ell}{\delta X} = 0.
                \end{equation}
    \end{enumerate}
\end{theorem}
For the proof we refer to \cite{holm1998euler}, where also the left/left, left/right, right/left versions of the theorem can be found. This theorem turns the ``Lagrangian plane" (left vertical plane) in Figure \ref{fig:cube} into a commutative diagram.

We have computed the expression for the $\mathrm{ad}^*$ operator in the previous section for the group of diffeomorphisms on the circle. Hence, given a right-invariant metric on $\mathfrak{D}^+$, we obtain the differential equation describing its geodesics from the Euler-Poincar\'e theorem. Given the $L^2$-metric \eqref{eq:l2metric}, one obtains by right-invariance the reduced Lagrangian. Expanding the vector field $X(x)=u(x)\partial_x$, we can express the reduced Lagrangian in terms of $u$ as follows
\begin{equation}
    \ell(u) = \frac{1}{2}\int_{S^1}u^2 dx.
\end{equation}
The functional derivative of the reduced Lagrangian is
\begin{equation}
    \frac{\delta\ell}{\delta u} = u,
\end{equation}
and the corresponding Euler-Poincar\'e equation is given by
\begin{equation}
    u_t + 3uu_x = 0,
\end{equation}
with the convention $\partial_x u = u_x$. This is recognised as the Hopf equation or, equivalently, as the inviscid Burgers equation. This equation is the prototypical example of a hyperbolic partial differential equation that produces shocks in finite time. Since the solutions of the geodesic equation describe the coadjoint orbits, here we directly see that the coadjoint orbits of $\mathfrak{D}^+$ are badly behaved for the $L^2$ metric. Next we compute the geodesic equation for the $H^1$-metric. This metric is given by 
\begin{equation}\label{eq:h1metric}
    L(\psi,\dot{\psi}) = \frac{1}{2}\int_{S^1} \alpha^2|\dot{\psi}|^2 + \beta^2|\dot{\psi}_x|^2 dx,
\end{equation}
where $\alpha,\beta\in\mathbb{R}$ are constants. This Lagrangian is again right-invariant under the tangent-lifted action of $\mathfrak{D}^+$, which allows us to compute the reduced Lagrangian. The reduced Lagrangian is given by
\begin{equation}\label{eq:varderh1}
    \ell(u) = \frac{1}{2}\int_{S^1} \alpha^2 u^2 + \beta^2 u_x^2 dx.
\end{equation}
The variational derivative can be computed to be
\begin{equation}
    \frac{\delta \ell}{\delta u} = \alpha^2 u-\beta^2 u_{xx}
\end{equation}
and the corresponding Euler-Poincar\'e equation is given by
\begin{equation}
    \alpha^2 u_t - \beta^2 u_{txx} + 3\alpha^2 uu_x - 2\beta^2 u_x u_{xx} - \beta^2 uu_{xxx} = 0.
\end{equation}
For $\alpha=\beta=1$, this is the famous Camassa-Holm equation, \cite{camassa1993integrable, camassa1994new}, which has applications in the modelling of water waves and image analysis. For $\alpha=1$ and $\beta = 0$ one obtains the Hopf equation as the geodesic equation. For $\alpha=0$ and $\beta = 1$, the $H^1$-norm becomes the homogeneous $H^1$-norm, denoted $\dot{H}^1$. The corresponding geodesic equation to the $\dot{H}^1$-metric is the Hunter-Saxton equation, \cite{hunter1991dynamics, hunter1994completely}, which appears in the context of the dynamics of director fields in nematic liquid crystals and is given by
\begin{equation}
    u_{txx} + 2u_x u_{xx} + uu_{xxx} = 0.
\end{equation}
Both the Camassa-Holm equation and the Hunter-Saxton equation are examples of completely integrable nonlinear partial differential equations, which in particular means that their behaviour is much better than that of the Hopf equation. The general solution behaviour is however quite different for the Camassa-Holm and Hunter-Saxton equation. The metric can improve the behaviour of the geodesic equation, but the metric is not the only means by which one can improve the geodesic equation. The development of numerical schemes for these equations is difficult in general, but the Euler-Poincar\'e framework provides guidance in the discretisation procedure. Namely, it is generally better to discretise the Euler-Poincar\'e equation \eqref{eq:eulerpoincare} and separately discretise the variational derivatives of the Lagrangian \eqref{eq:varderh1}. This avoids having to discretise the high-order derivatives that one encounters in the Camassa-Holm and Hunter-Saxton equations. However, due to the singular profiles of the soliton solutions to the Camassa-Holm and Hunter-Saxton equations, developing suitable numerical schemes is notoriously challenging, see \cite{chertock2012convergence, chertock2015elastic} for numerical schemes for the Camassa-Holm equation and \cite{holden2007convergent} for numerical schemes for the Hunter-Saxton equation. In the next section we discuss a topological improvement to $\mathfrak{D}^+$ by means of a central extension.

\section{Virasoro-Bott group}\label{sec:vir}
In this section we discuss the central extension of the Lie algebra of vector fields on the circle. This is the Lie algebra known as the Virasoro algebra. In infinite dimensions, not every Lie algebra is associated with a particular Lie group, but in the particular case of the Virasoro algebra, there is also an associated Lie group, known as the Virasoro-Bott group. We discuss Euler-Poincar\'e equations on the Virasoro-Bott group, which we call EPVir equations, in line with the convention in the previous section.

The Virasoro-Bott group is the universal central extension and the universal covering group for $\mathfrak{D}^+$. The Virasoro algebra is the universal central extension of $\mathfrak{X}(S^1)$. A central extension is the extension of a group or algebra by adding new elements in such a way that the new elements commute with all original elements. The purpose of central extensions is to create a larger group that has desired symmetries. It is of particular importance in quantum mechanics and string theory, where Bargmann's theorem states that if the second Lie algebra cohomology group $H^2(\mathfrak{g},\mathbb{R})$ is trivial, then every projective unitary representation of the Lie group $G$ can be replaced by ordinary unitary representations of the Lie group $G$ by passing to the universal covering group. For example, $\mathbb{R}^{2n}$ considered as a Lie algebra has nontrivial second cohomology, which means that Bargmann's theorem does not apply. However, $\mathbb{R}^{2n}$ has a universal central extension in the form of the Heisenberg algebra. The Heisenberg algebra has trivial second cohomology, so Bargmann's theorem applies. The unitary representations of the Heisenberg group are extensions of Fourier transform via the Stone-von Neumann theorem. For more details see \cite{bargmann1954unitary, hall2013quantum}. 

In the previous section we derived the EPDiff equations for various metrics and showed that the partial differential equation whose solution are geodesics with respect to the $L^2$-metric is the Hopf equation. This equation is known to produce shocks in finite time. By choosing different metrics, this issue can be resolved, which led to the Camassa-Holm and Hunter-Saxton equations. We show in this section that $\mathfrak{X}$ has nontrivial second cohomology. By a central extension, we can enlarge the algebra to obtain the Virasoro algebra, for which we can then compute the adjoint and coadjoint representations.

The space of vector fields $\mathfrak{X}$ is a Lie algebra whose bracket is given by
\begin{equation}
[u(x)\partial_x,v(x)\partial_x] = (uv_x - vu_x)\partial_x.
\end{equation}
The map $\omega:\mathfrak{X}(S^1)\times\mathfrak{X}(S^1)\to\mathbb{R}$, defined by
\begin{equation}
\omega\Big(u(x)\partial_x,v(x)\partial_x\Big) = \int_{S^1}u_x v_{xx}\,dx,
\end{equation}
is the Gel'fand-Fuchs 2-cocycle which extends the algebra $\mathfrak{X}(S^1)$ to the Virasoro algebra $\mathfrak{vir}$ in a nontrivial manner. The Gel'fand-Fuchs map is a 2-cocycle, which means that it satisfies the identity
\begin{equation}
\omega([u\partial_x,v\partial_x],w\partial_x) + \omega([w\partial_x,u\partial_x],v\partial_x) + \omega([v\partial_x,w\partial_x],u\partial_x) = 0,
\end{equation}
for $u\partial_x,v\partial_x,w\partial_x\in\mathfrak{X}$. The central extension of a Lie algebra is required to be a Lie algebra in its own right. The extended bracket must satisfy the Jacobi identity. The 2-cocycle identity above is necessary for the extended bracket to satisfy the Jacobi identity. We can then form the extended Lie bracket with $a,b\in\mathbb{R}$
\begin{equation}\label{eq:virbracket}
[(u\partial_x,a),(v\partial_x,b)]_{\mathfrak{vir}}=([u\partial_x,v\partial_x],\omega(u\partial_x,v\partial_x)),
\end{equation}
which defines the Virasoro algebra as $(\mathfrak{{vir}},[\,\cdot\,,\,\cdot\,]_{\mathfrak{vir}})=(\mathfrak{X}(S^1)\oplus\mathbb{R},([\,\cdot\,,\,\cdot\,],\omega(\,\cdot\,,\,\cdot\,))$. It can be shown that the Virasoro algebra is the unique (up to isomorphism) central extension of the algebra of vector fields on the circle. The proof of this statement can be found in \cite{khesin2008geometry}, which uses the fact that the second cohomology group $H^2(\mathfrak{X}(S^1),\mathbb{R})$ is one-dimensional and generated by the Gel'fand-Fuchs cocycle. It can also be shown that the Gel'fand-Fuchs cocycle is nontrivial, which means that it is not a coboundary. A coboundary is a map $\alpha:\mathfrak{X}(S^1)\to\mathbb{R}$ such that $\omega(X,Y) = \alpha([X,Y])$. In general, it is not guaranteed that there corresponds a group to centrally extended Lie algebra, but for the Virasoro algebra this is the case.

The Virasoro algebra corresponds to the group known as the Virasoro-Bott group, which is the central extension of the orientation preserving diffeomorphisms on the circle $\mathfrak{D}^+$ by the Bott-Thurston 2-cocycle $B:\mathfrak{D}^+\times\mathfrak{D}^+\to \mathbb{R}$. Let $\varphi,\psi\in \mathfrak{D}^+$, the Bott-Thurston 2-cocycle is then given by 
\begin{equation}
B(\varphi,\psi) = \frac{1}{2}\int_{S^1}\log(\varphi_x \circ\psi) d\log\psi_x,
\end{equation}
here given with normalisation constant $\frac{1}{2}$. The Bott-Thurston cocycle satisfies the group 2-cocycle identity
\begin{equation}
B(\varphi\circ\zeta,\psi)+B(\varphi,\zeta) = B(\varphi,\zeta\circ\psi)+B(\zeta,\psi).
\end{equation}
The group 2-cocycle identity is necessary to guarantee that the group operation on the Virasoro-Bott group is associative. The Bott-Thurston 2-cocycle is a continuous 2-cocycle on $\mathfrak{D}^+$ and the central extension is the Virasoro-Bott group $(Vir,\bullet) = (\mathfrak{D}^+\oplus\mathbb{R},(\circ,+))$, where the product is defined by
\begin{equation}
(\varphi,\alpha_1)\bullet(\psi,\alpha_2) = (\varphi\circ\psi,\alpha_1+\alpha_2+B(\varphi,\psi)).
\end{equation}
Note that the group action on the $\mathbb{R}$-part is abelian. Hence in the extended Lie bracket \eqref{eq:virbracket}, the contributions of $\alpha_1$ and $\alpha_2$ disappear. The Lie algebra corresponding to $Vir$ is the Virasoro algebra $\mathfrak{vir}$. By taking two one-parameter subgroups of the Virasoro-Bott group and differentiating them with respect to their parameters, we can obtain the adjoint and coadjoint representations.

The pairing on the Virasoro algebra is the sum of the pairing on $\mathfrak{X}$ and the Euclidean inner product on $\mathbb{R}$
\begin{equation}
    \langle (m dx\otimes dx, \varepsilon), (u\partial_x, a)\rangle_{\mathfrak{vir}} = \langle m dx\otimes dx, u\partial_x\rangle_{\mathfrak{X}} + \varepsilon a.
\end{equation}
With the pairing between the Virasoro algebra and its dual defined, we can compute the adjoint and coadjoint representations. For the Virasoro-Bott group and the Virasoro algebra, the adjoint and coadjoint operators can be computed following the strategy illustrated in Section \ref{sec:diff}. This method requires an exponential map, whose existence is shown in \cite{constantin2007geodesic}. We summarise the operations in Table \ref{tab:virasoro}.
\begin{table}[H]
\centering
\setlength{\extrarowheight}{6pt}
\begin{tabular}{ll}
$\mathrm{AD}_{(\varphi,\alpha)}:$ & $(\widetilde{\varphi},\widetilde{\alpha})\mapsto \big(\varphi\circ\widetilde{\varphi}\circ\varphi^{-1},\, \frac{1}{2} \int_{S^1}\log\big((\varphi\circ\widetilde{\varphi})_x\circ\varphi^{-1}\big)d(\log\varphi^{-1})_x + \widetilde{\alpha}\big)$\\
$\mathrm{Ad}_{(\varphi,\alpha)}:$ & $(u(x)\partial_x,a)\mapsto\big(u(\varphi)\partial_{\varphi},\, a+\frac{1}{2}\int_{S^1}(u\circ\varphi^{-1})d\log(\varphi^{-1})_x\big)$\\
$\mathrm{Ad}^*_{(\varphi,\alpha)^{-1}}:$ & $\big(m(x)dx\otimes dx,\varepsilon\big)\mapsto \Big(\big(m(\varphi)(\varphi')^2 + a(S\varphi)(x)\big)dx\otimes dx,\, a\Big)$\\
$\mathrm{ad}_{(u\partial_x,a)}:$ & $(v\partial_x,b)\mapsto \Big((uv_x-vu_x)\partial_x,\, \int_{S^1}u_x v_{xx} dx\Big)$\\
$\mathrm{ad}^*_{(u\partial_x,a)}:$ & $\big(m(x)dx\otimes dx,\varepsilon\big)\mapsto ((mu)_x + mu_x+ \varepsilon u_{xxx})dx\otimes dx,\, 0\big)$
\end{tabular}
\caption{The adjoint and coadjoint actions and operators of the Virasoro-Bott group and the Virasoro algebra.}
\label{tab:virasoro}
\end{table}
The central extension variable influences the momentum variable but plays no significant role in the coadjoint actions ${\rm Ad}^*$ and ${\rm ad}^*$. This automatically implies that the variable $\varepsilon$ in the pair $(m(x)dx\otimes dx, \varepsilon)$ has no dynamics of itself and can thus be interpreted as a parameter. In the coadjoint action ${\rm Ad}^*$, it should be noted that $m(\varphi)(\varphi')^2(dx\otimes dx) = m(\varphi)(d\varphi \otimes d\varphi)$. Furthermore, in the coadjoint action of the group, the operator $(S\varphi)(x)$ is called the Schwarzian derivative of the diffeomorphism $\varphi$, which is given by
\begin{equation}
(S\varphi)(x) = \frac{\varphi_{xxx}}{\varphi_x}-\frac{3}{2}\left(\frac{\varphi_{xx}}{\varphi_{x}}\right)^2.
\end{equation}
The Schwarzian derivative is an operator that appears in various areas of mathematics as shown by \cite{ovsienko2009schwarzian} and can be shown to be group 1-cocycle in the sense of \cite{souriau1970structure}. It is invariant under M\"obius transformations, i.e, for $a,b,c,d\in\mathbb{R}$, $\varphi,\psi\in \mathfrak{D}(S^1)$ and $x\in S^1$, we have
\begin{equation}
(S\varphi)(x) = (S\psi)(x)\quad \text{ implies }\quad \psi(x) = \frac{a\varphi(x) + b}{c\varphi(x) + d}.
\end{equation} 
The invariance under $SL(2,\mathbb{R})$ links the Schwarzian derivative with the theory of modular forms and is the primary reason for the characterisation of coadjoint orbits of the Virasoro-Bott group being better than the coadjoint orbits of the diffeomorphism group over the circle.

Now we can take the $H^1$-metric as in \eqref{eq:h1metric}, the variational derivative as in \eqref{eq:varderh1} and obtain the corresponding Euler-Poincar\'e equation on the Virasoro-Bott group 
\begin{equation}
\begin{aligned}
    \alpha^2 u_t - \beta^2 u_{txx} + 3\alpha^2 uu_x - 2\beta^2 u_x u_{xx} - \beta^2 uu_{xxx} + \varepsilon u_{xxx} &= 0,\\
    \varepsilon_t &= 0.
\end{aligned}
\end{equation}
For $\alpha=1$ and $\beta=0$, we obtain the Korteweg-De Vries equation. For $\alpha=1$ and $\beta=1$, we obtain the dispersive Camassa-Holm equation and for $\alpha=0$ and $\beta=1$, we obtain the dispersive Hunter-Saxton equation. In summary, we have the following Euler-Poincar\'e equations of geodesic type on $\mathfrak{D}^+$ and $\mathrm{Vir}$ in Table \ref{tab:equations}.

\begin{table}[H]
    \centering
    \setlength{\extrarowheight}{6pt}
    \begin{tabular}{c|c|r}
        Group & Metric & Equation \\
        \hline $\mathfrak{D}^+$ & $L^2$ & $u_t + 3uu_x = 0$ \\
        $\mathfrak{D}^+$ & $H^1$ & $u_t - u_{txx} + 3uu_x - 2u_x u_{xx} - uu_{xxx}  = 0$ \\
        $\mathfrak{D}^+$ & $\dot{H}^1$ & $u_{txx} + 2u_x u_{xx} + uu_{xxx} = 0$ \\
        $\mathrm{Vir}$ & $L^2$  & $u_t + 3uu_x + \varepsilon u_{xxx} = 0$ \\
        $\mathrm{Vir}$ & $H^1$ & $u_t - u_{txx} + 3uu_x - 2u_x u_{xx} - uu_{xxx} + \varepsilon u_{xxx} = 0$ \\
        $\mathrm{Vir}$ & $\dot{H}^1$ & $u_{txx} + 2u_x u_{xx} + uu_{xxx} - \varepsilon u_{xxx} = 0$
    \end{tabular}
    \caption{Geodesic equations for various combinations of group and metric.}
    \label{tab:equations}
\end{table}

Having derived the geodesic equations for various metrics on the group of orientation-preserving diffeomorphisms over the circle and on the Virasoro-Bott group, let us show that they are indeed geodesic equations and investigate associated local wellposedness questions.

\section{Geometric analysis of geodesic equations}
Based on the results of \cite{arnold1966geometrie} from the geometric point of view, a few years later \cite{ebin1970groups} introduced rigourous techniques to study the wellposedness of geodesic equations on diffeomorphism groups. These rigourous techniques require to view the diffeomorphism group as a Hilbert or Banach manifold and imply local existence and uniqueness of strong solutions in spaces of high regularity. From the point of view of analysis of partial differential equations, the necessary regularity for these methods to apply is often considered too high. Hence, there are many papers that investigate the wellposedness questions of the PDEs in Table \ref{tab:equations} in much weaker settings, but this requires a case by case approach. The geometric techniques can tackle four of the six PDEs at once. 

In the Fr\'echet setting, while both the left and the right action are smooth, we do not have access to fixed point iteration. This is problematic, since we cannot directly apply results from the theory of ordinary differential equations. So we sacrifice the idea that the diffeomorphism group is a Lie group in the Fr\'echet sense and deal with it as a topological group instead. For diffeomorphism groups from the topological viewpoint, the right action is smooth, but the left action is only continuous. Upon agreeing to only use right actions, we can enjoy both the geometric and the analytical techniques. The results in this section can be found with more details in \cite{shkoller1998geometry}. For further discussion see also \cite{ovsienko1987korteweg, shkoller1998geometry, misiolek1998shallow, kouranbaeva1999camassa, lenells2007hunter, michor2007some, lenells2008hunter}.

We enlarge the space of smooth diffeomorphisms to the space $\mathrm{Diff}^s(S^1) = \{f\in H^s(S^1,S^1)\,\vert\, f\text{ bijective and } f^{-1}\in H^s(S^1,S^1)\}$ with the requirement that $s>3/2$ in order for $\mathrm{Diff}^s(S^1)$ to be open in $H^s(S^1,S^1)$ to qualify as a Banach manifold. The corresponding algebra $\mathfrak{X}^s(S^1)$ is then a Banach algebra that can be identified with the Sobolev space $H^s(S^1)$. For the analysis of the equations, we can use the fact that we derived the equations using the Euler-Poincar\'e theorem. Specifically, we use the reconstruction equation $\psi_t = u\circ\psi$ in combination with the PDEs. Note that all PDEs can be formulated in the same manner by virtue of being Euler-Poincar\'e equations. On $\mathrm{Diff}^s(S^1)$ we have 
\begin{equation}
    Au_t + uAu_x + 2u_x Au = 0
\end{equation}
where the operator $A=\alpha^2 - \beta^2\partial_x^2$. Since $A:H^s\to H^{s-2}$ for $\alpha,\beta\neq 0$, a quick count of derivatives shows that the second term in the PDE above requires an extra derivative compared to the other terms. Upon formulating the Virasoro-Bott group as a Banach manifold in the same manner as for the diffeomorphisms, we can write on $\mathrm{Vir}^s$
\begin{equation}
    Au_t + uAu_x + 2u_x Au + \varepsilon u_{xxx} = 0,
\end{equation}
with $A=\alpha^2 - \beta^2\partial_x^2$ as before and $\varepsilon>0$. Since the operator $A$ is invertible on $H^s(S^1)$ if $s\geq 2$, its inverse $A^{-1}:H^{s-2}\to H^s$. Note that for $\beta=0$ the operator $A$ reduces to a scaling operator, which means that its inverse does not induce any smoothing. Let us focus first on geodesics on the group of diffeomorphisms over the circle. The geodesic equation can be obtained by differentiating the reconstruction equation
\begin{equation}
\begin{aligned}
    \psi_{tt} &= u_t\circ\psi + (u_x\circ\psi)\psi_t\\
    &= (u_t + uu_x)\circ\psi.
\end{aligned}
\end{equation}
Now we substitute in the Euler-Poincar\'e equation. This yields
\begin{equation}
    \psi_{tt} = \Big(A^{-1}\big(-uAu_x - 2u_x Au + A(uu_x)\big)\Big)\circ\psi.
\end{equation}
Expanding $A(uu_x)$ produces a term that cancels the highest order derivative. To be precise $A(uu_x) = uA u_x + \sum_{i=0}^k (-1)^i \sum_{j=1}^{2i} {2i\choose j} (\partial_x^j u)(\partial_x^{2i-j+1} u) =: uAu_x + B(u,u)$. So this means that we are left with a bilinear form that we call $B(u,u)$. This bilinear form is obtained from the right-invariant metric (here denoted $\langle\,\cdot\,,\,\cdot\,\rangle$ on $\mathrm{Diff}^s(S^1)$ and the Lie bracket on $\mathfrak{X}^s(S^1)$ in the following manner
\begin{equation}
    \langle B(w,u),v\rangle = \langle w, [u,v]\rangle.
\end{equation}
Note that the bilinear form we obtain from the derivation above is a map $B:H^s(S^1)\times H^s(S^1)\to H^{s-2}(S^1)$. The geodesic equation now takes the form
\begin{equation}
    \psi_{tt} = \Big(A^{-1}\big(B(u,u)\big)\Big)\circ\psi.
\end{equation}
If the right hand side of the geodesic equation were a function of $\psi_t,\psi$, the proof for local wellposedness would be complete in $H^s(S^1)$. However, since $u=\psi_t\circ\psi^{-1}$, so this argument does not yet apply. Given a map $P:H^s(S^1)\to H^{s-2}(S^1)$, the related map $\tilde{P}_{\psi}:T_\psi\mathrm{Diff}^s(S^1)\to T_\psi^{s-2}\mathrm{Diff}^s(S^1)$ can be explicitly constructed by 
\begin{equation}
    \tilde{P}_\psi V = (P(V\circ\psi^{-1}))\circ\psi.
\end{equation}
Applying this to the geodesic equation yields
\begin{equation}
    \psi_{tt} = \tilde{A}_{\psi}^{-1}\big(\tilde{B}_{\psi}(\psi_t,\psi_t)\big).
\end{equation}
To conclude local wellposedness of the Camassa-Holm and Hunter-Saxton equations in $H^2(S^1)$, it remains to show that the map $T\mathrm{Diff}^s(S^1)\to T^{s-1}\mathrm{Diff}^s(S^1)$ given by $(\psi_t,\psi)\mapsto (\partial_x(\psi_t\circ\psi^{-1}))\circ\psi$ is smooth. This is verified with a quick computation that shows the naturality of the logarithmic derivative in these geodesic problems
\begin{equation}
    \begin{aligned}
        (\partial_x(\psi_t\circ\psi^{-1}))\circ \psi &= \psi_{tx}((\partial_x \psi^{-1})\circ\psi)\\
        &= \frac{\psi_{tx}}{\psi_x}\\
        &= \partial_t \log(\psi_x).
    \end{aligned}
\end{equation}
The logarithmic derivative shows that $(\psi_t,\psi)\mapsto (\partial_x(\psi_t\circ\psi^{-1}))\circ\psi$ is a smooth map. This means that the geodesic spray is smooth and by theorems of ordinary differential equations on Banach spaces, we can conclude local wellposedness of the Camassa-Holm and Hunter-Saxton equations in $H^s(S^1)$. This argument does not apply to the Hopf equation, since the $L^2$-metric does not introduce smoothing. It can be further established that the Euler-Poincar\'e equations on the group of diffeomorphisms with $H^s$-metrics with $s>3/2$ depend on initial data in a $C^1$ manner.

For the Virasoro-Bott group, a similar approach can be used, as shown in detail in \cite{michor2007some}. The conclusion here is that there exists a unique $C^3$-geodesic for the right-invariant $H^k$-metric with $k\geq 2$ and the solution depends in a $C^1$-manner on the initial data. For the details we refer to \cite{michor2007some}, but let us sketch the idea. We enlarge the smooth Virasoro-Bott group $\mathrm{Vir}$ to the Virasoro-Bott group $\mathrm{Vir}^s$ with the Sobolev topology $H^s$ with $s>3/2$. The Euler-Poincar\'e equation on $\mathrm{Vir}^s$ with the operator $A= \alpha^2 - \beta^2\partial_x^2$ 
\begin{equation}
    Au_t + uAu_x + 2u_x Au + \varepsilon u_{xxx} = 0. 
\end{equation}
Manipulating the expression as before, we derive
\begin{equation}
    \psi_{tt} = \Big(A^{-1}(-uA u_x-2u_x Au + A(uu_x) - \varepsilon u_{xxx})\Big)\circ\psi.
\end{equation}
Defining the operator $C:\mathfrak{vir}^s \to \mathfrak{vir}^{s-2}$ 
\begin{equation}
    C(u,\varepsilon) = -2u_x Au + B(u,u) -\varepsilon u_{xxx},
\end{equation}
we obtain 
\begin{equation}
    \psi_{tt} = \Big(A^{-1}\big(C(u,\varepsilon)\big)\Big)\circ\psi.
\end{equation}
It is now a matter of showing that we can formulate the above equation strictly in terms of variables on $\mathrm{Vir}^s$. For this part, we refer to \cite{michor2007some}. The geodesic spray on the Virasoro-Bott group for the $H^k$ metric with $k\geq 2$ is then shown to be smooth and one can proceed to show that there is $C^1$-dependence on initial data. Next we analyse the symmetries of the equations themselves. For this, we use the theory of infinitesimal symmetry analysis in the next section.

\section{Symmetry groups of differential equations}\label{sec:symmetry}
The symmetry group of a differential equation is the largest local group of transformations acting on the independent and dependent variables of the system with the property that it transforms (smooth) solutions of the system to other (smooth) solutions. In \cite{olver1993applications}, one can find a systematic, computational method to explicitly describe the symmetry group of any given system of differential equations. Obtaining any explicit solution is of great interest since they can be used as models for physical experiments or as benchmarks for testing numerical methods, but this is in general very challenging. While the symmetry groups do not provide explicit solutions, they can be used to test numerical methods. One of the main advantages of knowing a symmetry group is that one can construct new solutions from known ones. In this section, we employ the method described by \cite{olver1993applications} to obtain the symmetry groups of the nonlinear partial differential equations in Table \ref{tab:equations}. 

The method relies on the notion of the $n$-th order jet space $X\times U^{(n)}$ whose coordinates represent the independent variables, dependent variables and the derivatives up to order $n$ of the dependent variables. Given a system of $l$ $n$-th order differential equations in $p$ independent and $q$ dependent variables, the underlying solution space can be thought of as the graph $X\times U$. The $n$-th order prolongation $\mathrm{pr}^{(n)}$ maps the graph $X\times U$ to the $n$-th order jet space $X\times U^{(n)}$.

Let $\mathfrak{S}$ be a system of (partial) differential equations. A symmetry group of the system $\mathfrak{S}$ is a local group $G$ of transformations acting on an open subset $M$ of the space of independent and dependent variables for the system with the property that whenever $u=f(x)$ is a (smooth) solution of $\mathfrak{S}$, and whenever $g\cdot f$ is defined for $g\in G$, then $u=g\cdot f(x)$ is also a (smooth) solution of the system.

A system $\mathfrak{S}$ of $l$ $n$-th order differential equations in $p$ independent and $q$ dependent variables is given as a system of equations
\begin{equation}\label{eq:system}
    \Delta_v(x,u^{(n)}) = 0, \quad v=1,\hdots,l,
\end{equation}
where $x=(x^1,\hdots,x^p)$, $u=(u^1,\hdots, u^q)$ and the derivatives of $u$ with respect to $x$ up to order $n$. In our case, for the equations in Table \ref{tab:equations}, we have $x=(t,x)$ and $u=u$. The functions $\Delta(x,u^{(n)}) = (\Delta_1(x,u^{(n)}),\hdots, \Delta_l(x,u^{(n)}))$ are assumed to be smooth in their arguments. This allows $\Delta$ to be viewed as a smooth map from the jet space $X\times U^{(n)}$ to an $l$-dimensional Euclidean space. The differential equations determine a subvariety 
\begin{equation}
    \mathfrak{S}_\Delta = \{(x,u^{(n)}):\Delta(x,u^{(n)})=0\}\subset X\times U^{(n)}
\end{equation}
of the total jet space. A symmetry of this subvariety is described in terms of the prolonged group action $\mathrm{pr}^{(n)}G$. The system \eqref{eq:system} is said to be of maximal rank if the $l\times(p+qp^{(n)})$ Jacobian matrix
\begin{equation}
    J_\Delta(x,u^{(n)}) = \left(\frac{\partial \Delta_v}{\partial x^i},\frac{\partial\Delta_v}{\partial u_J^\alpha}\right)
\end{equation}
of $\Delta$ with respect to all the variables $(x,u^{(n)})$ is of rank $l$ whenever $\Delta(x,u^{(n)})=0$. The maximal rank condition is a necessary condition for the following theorem. 

\begin{theorem}
    Let $\Delta:X\times U^{(n)}\to\mathbb{R}^l$ define a system of differential equations $\Delta_v(x,u^{(n)})=0$, $v=1,\hdots,l$, on $M\subset X\times U$ of maximal rank. Then $G$ is a symmetry group of the system if and only if
    \begin{equation}\label{eq:invariancecriterion}
        \mathrm{pr}^{(n)}\mathbf{v}[\Delta_v(x,u^{(n)})] = 0, \quad v=1,\hdots,l,\quad \text{whenever} \quad \Delta(x,u^{(n)})=0,
    \end{equation}
    for every infinitesimal generator $\mathbf{v}$ of $G$.
\end{theorem}

The proof of this theorem can be found in chapter 2 of \cite{olver1993applications}. 

The computational procedure to obtain the symmetry groups of the partial differential equations in Table \ref{tab:equations} is as follows. The jet spaces $X\times U^{(n)}$ are defined over the base space $\mathbb{R}^2\times\mathbb{R}$, since we have two independent variables $t,x$ denoting time and space, and we have a single dependent variable $u$. We let the coefficients $X(t,x,u)$, $T(t,x,u)$ and $U(t,x,u)$ of the infinitesimal generator $\mathbf{v}$ of a hypothetical one-parameter symmetry group be unknown functions of $t,x,u$. The coefficients of the prolonged infinitesimal generator $\mathrm{pr}^{(n)}\mathbf{v}$ will be certain explicit expressions that involve the partial derivatives of $u$ with respect to $t,x$ and $u$. The infinitesimal criterion of invariance \eqref{eq:invariancecriterion} involves $t, x, u$, the derivatives of $u$ with respect to $t$ and $x$, and $X(t,x,u), T(t,x,u), U(t,x,u)$ and all their partial derivatives. After the elimination of any dependencies among the derivatives of $u$ caused by the system itself, it is then possible to equate the coefficients of the remaining unconstrained partial derivatives of $u$ to zero. The result is a large number of simple partial differential equations for the coefficient functions $X,T$ and $U$ of the infinitesimal generator. The solution to these simple PDEs determines the most general infinitesimal symmetry of the system and the system of infinitesimal generators forms a Lie algebra of symmetries. The symmetry group is obtained by exponentiation of the obtained vector fields.

Given the equations in Table \ref{tab:equations}, we find the equation for which the vector field $\mathbf{v}=T\partial_t + X\partial_x + U\partial_u$ generates a one-parameter symmetry group. We obtain the following equations for the coefficients of the infinitesimal generator of a hypothetical one-parameter symmetry group
\begin{table}[H]
    \centering
    \setlength{\extrarowheight}{6pt}
    \begin{tabular}{l|l}
    Equation &  Coefficients\\
    \hline
    Hopf     & $T(t,x,u)\, =\frac{1}{2} F(u, -tu + x)t^2 + F(u, -tu + x)t + H_1(-tu+x)$ \\
    $u_t + u u_x = 0$ & $X(t,x,u) =  -\frac{1}{2} F(u, -tu + x)t^2 u + (F(u, -tu + x)u$ \\
    & $\qquad \qquad \qquad + F(u,-tu+x))t + H_2(-tu+x)$ \\
     & $U(t,x,u)\, = F(u, -tu + x)$ \\
    \hline
    Camassa-Holm  & $T(t,x,u) =  -c_1 t +c_2$ \\
    $u_t - u_{txx} + 3u u_x = 2u_x u_{xx} + u u_{xxx}$ & $X(t,x,u) =  c_3$   \\
    & $U(t,x,u) = c_1u$ \\
    \hline
    Hunter-Saxton & $T(t,x,u) =  c_3F_1(t) - c_1t + c_2$ \\
    $u_{txx} +  2u_x u_{xx}  + u u_{xxx} =0 $ & $X(t,x,u) =  c_3\frac{\partial F_1(t)}{\partial t}x + c_4F_2(t)$   \\
    & $U(t,x,u) = c_1u + c_3\frac{\partial^2 F_1(t)}{\partial t^2}x +  c_4\frac{\partial F_2(t)}{\partial t}$ \\
    \hline
    Korteweg-De Vries & $T(t,x,u) =  c_2 + 3c_4t$ \\
    $u_t + 3u u_x + u_{xxx} = 0$ & $X(t,x,u) =  c_1 + c_3t + c_4x$ \\
    & $U(t,x,u) = -2c_4u + c_3$ \\
    \hline
    Dispersive Camassa-Holm & $T(t,x,u) =  -c_1 t +c_2$ \\
    $u_t - u_{txx} + 3u u_x + u_{xxx} = 2u_x u_{xx} + u u_{xxx} $ & $X(t,x,u) =  \frac{3}{2}c_1 t + c_3$   \\
    & $U(t,x,u) = c_1 u + \frac{1}{2} c_1$ \\
    \hline
    Dispersive Hunter-Saxton & $T(t,x,u) =  c_3F_1(t) - c_1t + c_2$ \\
    $u_{txx} +  2u_x u_{xx}  + u u_{xxx} - u_{xxx}=0$ & $X(t,x,u) =  c_3\frac{\partial F_1(t)}{\partial t}x + c_4F_2(t)$   \\
    & $U(t,x,u) = c_1u + c_3\frac{\partial^2 F_1(t)}{\partial t^2}x +  c_4\frac{\partial F_2(t)}{\partial t} + c_1$
    \end{tabular}
    \caption{The coefficients of the infinitesimal generator. Note that the coefficients of the Hopf equation and the two Hunter-Saxton equations are function valued.}
    \label{tab:coefficients}
\end{table}
In Table \ref{tab:coefficients} it can be noted that the coefficients for the Hopf equation are given implicitly and they are highly nonlinear. For this reason, we do not include the Hopf equation in our further symmetry analysis. For the other equations the coefficients are much simpler. While the Hunter-Saxton equation also features the arbitrary functions $F_1(t)$ and $F_2(t)$, these fortunately only depend on time. From the coefficients, we compute the vector fields that span the symmetry algebra, given in Table \ref{tab:spanning}. 

\begin{table}[H]
    \centering
    \setlength{\extrarowheight}{6pt}
    \begin{tabular}{l|l}
        Camassa-Holm equation \\
        \hline
        $\mathbf{v}_1 = -t\partial_t + u\partial_u $ & \quad \text{scaling}\\
        $\mathbf{v}_2 = \partial_t$ & \quad \text{time translation} \\
        $\mathbf{v}_3 = \partial_x $ & \quad \text{space translation} \\
        \hline \\
        Hunter-Saxton equation \\
        \hline
        $\mathbf{v}_1 = -t\partial_t + u\partial_u$ & \quad \text{scaling}\\
        $\mathbf{v}_2 = \partial_t$ & \quad \text{time translation} \\
        $\mathbf{v}_3 = F_1(t)\partial_t + x\frac{\partial F_1(t)}{\partial t}\partial_x + x\frac{\partial^2 F_1(t)}{\partial t^2}\partial_u$ & \quad \text{generalised scaling} \\
        $\mathbf{v}_4 = F_2(t)\partial_x + \frac{\partial F_2(t)}{\partial t}\partial_u $ & \quad \text{generalised Galilean boost} \\
        \hline \\
        Korteweg-De Vries equation \\
        \hline
        $\mathbf{v}_1 = \partial_x$ & \quad \text{space translation}\\
        $\mathbf{v}_2 = \partial_t$ & \quad \text{time translation} \\
        $\mathbf{v}_3 = t\partial_x + \partial_u$ & \quad \text{Galilean boost} \\
        $\mathbf{v}_4 = x\partial_x + 3t \partial_t -2u\partial_u$ & \quad \text{scaling} \\
        \hline \\
        Dispersive Camassa-Holm equation \\
        \hline
        $\mathbf{v}_1 = -t\partial_t + (u+\frac{1}{2})\partial_u + \frac{3}{2}t\partial_x $ & \quad \text{scaling}\\
        $\mathbf{v}_2 = \partial_t$ & \quad \text{time translation} \\
        $\mathbf{v}_3 = \partial_x $ & \quad \text{space translation} \\
        \hline \\
        Dispersive Hunter-Saxton equation \\
        \hline
        $\mathbf{v}_1 = -t\partial_t + (u+1)\partial_u$ & \quad \text{scaling}\\
        $\mathbf{v}_2 = \partial_t$ & \quad \text{time translation} \\
        $\mathbf{v}_3 = F_1(t)\partial_t + x\frac{\partial F_1(t)}{\partial t}\partial_x + x\frac{\partial^2 F_1(t)}{\partial t^2}\partial_u$ & \quad \text{generalised scaling} \\
        $\mathbf{v}_4 = F_2(t)\partial_x + \frac{\partial F_2(t)}{\partial t}\partial_u $ & \quad \text{generalised Galilean boost} \\
    \end{tabular}
    \caption{Spanning vector fields for symmetry algebras.}
    \label{tab:spanning}
\end{table}
In Table \ref{tab:spanning}, for the two types of Hunter-Saxton equation, we call the scaling symmetry a generalised scaling symmetry because the coefficients in the vector fields $\mathbf{v}_3$ are potentially nonlinear. Similarly, we call $\mathbf{v}_4$ for the Hunter-Saxton equations a generalised Galilean boost, because the coefficients are allowed to be arbitrary functions of time.

In order for the vector fields to span the corresponding symmetry algebra, their commutator table should be closed. While this is not an issue for the Korteweg-De Vries and Camassa-Holm equations, but for general choices of $F_1$ and $F_2$ the vector fields of the Hunter-Saxton equations do not span a symmetry algebra because their commutator tables are not closed. However, it can be shown algebraically that the only obtainable choices for $F_1$ and $F_2$ to close their commutator tables are: $F_1 = t + c$ and $F_2 = c$, the constant $c\in\mathbb{R}$ being the same for $F_1$ and $F_2$. Note that the coefficient for $t$ in $F_1$ must be one in order for the commutator tables to close. The new vector fields $\mathbf{v}_3$ and $\mathbf{v}_4$ for the Hunter-Saxton equations then take the forms $\mathbf{v}_3 = (t+1)\partial_t + x\partial_x$ and $\mathbf{v}_4 = \partial_x$, with $c=1$. The vector field $\mathbf{v}_3$ becomes another scaling symmetry and $\mathbf{v}_4$ is now recognised as space translation, which also could have been deduced from the fact that the Hunter-Saxton equations have solitary solutions, confirming the choice of $F_2 = c$.

It can now be noted that all equations have scaling, time translation and space translation symmetries. This is consistent with their solitary solutions. The Korteweg-De Vries equation is invariant under Galilean transformations. The Camassa-Holm equation is not Galilean invariant, so its symmetry algebra is lower dimensional than for the Korteweg-De Vries equation. With the choices for $F_1$ and $F_2$, the Hunter-Saxton equation and its dispersive version are also not Galilean invariant, however, the generalised Galilean boosts in Table \ref{tab:spanning} reduce to a new scaling symmetry. Hence, the symmetry algebra of the Hunter-Saxton equations remains spanned by four independent vector fields.

By exponentiating the spanning vector fields, we obtain the one-parameter groups $G_i$. The entries $G_i$ correspond to the transformed point $\exp(
\epsilon \mathbf{v}_i)(t,x,u) = (\tilde{t},\tilde{x},\tilde{u})$. Since each group $G_i$ is a symmetry group, this implies that if $u=f(t,x)$ is a solution, then so are the $u^{(i)}$ given in Table \ref{tab:sol}.
    
\begin{table}[H]
    \centering
    \setlength{\extrarowheight}{6pt}
    \begin{tabular}{l|l|l}
        Camassa-Holm equation & &\\
        \hline
        $\mathbf{v}_1 = -t\partial_t + u\partial_u $ & \, $G_1: (te^{-\epsilon},x, ue^\epsilon)$ & $ u^{(1)} = e^\epsilon f(te^{\epsilon},x)$\\
        $\mathbf{v}_2 = \partial_t$ & \, $G_2: (t+\epsilon,x, u)$ & $ u^{(2}) = f(t -\epsilon,x)$ \\
        $\mathbf{v}_3 = \partial_x $ & \, $G_3: (t,x+\epsilon, u)$ & $ u^{(3)} = f(t,x-\epsilon)$\\
        \hline & & \\
        Hunter-Saxton equation with & &\\ $F_1(t)=t+1$ and $F_2(t)=c=1$  & & \\
        \hline
        $\mathbf{v}_1 = -t\partial_t + u\partial_u$ & \, $G_1: (te^{-\epsilon},x, ue^\epsilon)$ & $ u^{(1)} = e^\epsilon f(te^{\epsilon},x)$\\
        $\mathbf{v}_2 = \partial_t$ & \, $G_2: (t+\epsilon,x, u)$ & $ u^{(2)} = f(t -\epsilon,x)$\\
        $\mathbf{v}_3 = (t+1)\partial_t + x\partial_x$ & \, $G_3: ((t+1)e^{\epsilon}, xe^\epsilon, u)$ & $ u^{(3)} = f((t-1)e^{-\epsilon},x^{-\epsilon})$ \\
        $\mathbf{v}_4 = \partial_x  $ & \, $G_4: (t,x+\epsilon, u)$ & $ u^{(4)} = f(t,x-\epsilon)$\\
        \hline & & \\
        Korteweg-De Vries equation & &\\
        \hline 
        $\mathbf{v}_1 = \partial_x$ & \, $G_1: (t,x+\epsilon, u)$ & $ u^{(1)} = f(t,x-\epsilon)$\\
        $\mathbf{v}_2 = \partial_t$ & \, $G_2: (t+\epsilon,x, u)$ & $ u^{(2)} = f(t-\epsilon,x)$ \\
        $\mathbf{v}_3 = t\partial_x + \partial_u$ & \, $G_3: (t,x+\epsilon t, u+\epsilon)$ & $ u^{(3)} = f(t, x-\epsilon) +\epsilon$  \\
        $\mathbf{v}_4 = x\partial_x + 3t \partial_t -2u\partial_u$ & \, $G_4: (te^{3\epsilon}, xe^\epsilon, u^{-2\epsilon})$ & $ u^{(4)} = e^{-2\epsilon} f(te^{-3\epsilon}, xe^{-\epsilon})$\\
        \hline & & \\
        Dispersive Camassa-Holm equation & & \\
        \hline
        $\mathbf{v}_1 = -t\partial_t + (u+\frac{1}{2})\partial_u + \frac{3}{2}t\partial_x $ & \, $G_1: (te^{-\epsilon},x+\frac{3}{2}t(1-e^{-\epsilon}), (u+\frac{1}{2})e^\epsilon - \frac{1}{2})$ & $ u^{(1)} = e^\epsilon f(te^{\epsilon},x+\frac{3}{2}t(1-e^{\epsilon}))$ \\
        & & $\qquad \quad +\frac{1}{2}(e^\epsilon - 1)$\\
        $\mathbf{v}_2 = \partial_t$ & \, $G_2: (t+\epsilon,x, u)$ & $ u^{(2)} = f(t -\epsilon,x)$\\
        $\mathbf{v}_3 = \partial_x $ & \, $G_3: (t,x+\epsilon, u)$ & $ u^{(3)} = f(t,x-\epsilon)$\\
        \hline & &\\
        Dispersive Hunter-Saxton equation & & \\ with $F_1(t)=t+1$ and $F_2(t)=c=1$ & & \\
        \hline
        $\mathbf{v}_1 = -t\partial_t + (u+1)\partial_u$ & \, $G_1: (te^{-\epsilon},x, (u+1)e^\epsilon-1)$ & $ u^{(1)} = e^\epsilon f(te^{\epsilon},x) +e^\epsilon -1$\\
        $\mathbf{v}_2 = \partial_t$ & \, $G_2: (t+\epsilon,x, u)$ & $ u^{(2)} = f(t -\epsilon,x)$\\
        $\mathbf{v}_3 = (t+1)\partial_t + x\partial_x$ & \, $G_3: ((t+1)e^{\epsilon}, xe^\epsilon, u)$ & $ u^{(3)} = f((t-1)e^{-\epsilon},x^{-\epsilon})$\\
        $\mathbf{v}_4 = \partial_x $ & \, $G_4: (t,x+\epsilon, u)$ & $ u^{(4)} = f(t,x-\epsilon)$ \\
    \end{tabular}
    \caption{In the first column we recall the vector fields that span the symmetry algebra. The second column shows the one-parameter groups corresponding to the vector fields and the third column shows how a solution is transformed by the symmetry group to another solution.}
    \label{tab:sol}
\end{table}

Table \ref{tab:sol} shows how solutions to the equations listed in Table \ref{tab:equations} transform to other solutions. These results can be used to benchmark and test numerical schemes for the equations. The coadjoint orbits of the Virasoro-Bott group for the $L^2$-metric behave much better than the coadjoint orbits of the diffeomorphism group for the $L^2$-metric, which is also noticeable in the symmetry analysis. For the $H^1$- and $\dot{H}^1$-metric, the influence of the central extension is not more than a change in the scaling group for solutions.

\section{Discussion and conclusion}\label{sec:conclusion}
In this work we derived and performed symmetry analysis of several geodesic equations on the group of diffeomorphisms on the circle and the Virasoro-Bott group. After introducing the preliminary topological and geometric notions, we showed that several important and famous equations in water waves and liquid crystals arise as geodesic equations on infinite-dimensional groups and that their solutions live on coadjoint orbits, by deriving them using the Euler-Poincar\'e theorem. This derivation makes explicit use of the right-invariance of the metrics under the action of the diffeomorphism group over the circle or the Virasoro-Bott group. In particular, we show that the universal central extension of the diffeomorphism group over the circle is able to improve the behaviour of the coadjoint orbits for the $L^2$-metric. Importantly, this derivation provides guidance on the numerical discretisation of such equations. Indeed, the Euler-Poincar\'e equation together with the variational derivatives of the Lagrangian can be viewed as a coupled system of partial differential equations that can be solved successively, rather than as a single partial differential equation with high order derivatives. However, obtaining suitable numerical discretisations of these equations is challenging in general. This is why we compute the symmetry groups of the equation to aid the benchmarking and testing of potential numerical schemes.

Table \ref{tab:sol} shows how solutions to the equations listed in Table \ref{tab:equations} transform to other solutions. These symmetry groups allow us to investigate the influence of the central extension on metrics associated with better-behaved geodesic equations. To illustrate once more the poor behaviour of the Hopf equation, we showed that the symmetry group of the Hopf equation is much more complicated than for any of the other equations. For the Camassa-Holm and Hunter-Saxton equations, the transition from the diffeomorphism group to its universal central extension changes their scaling symmetry group, but leaves their other symmetry groups invariant. For the Hunter-Saxton equations, the symmetry groups feature arbitrary functions of time, which in general do not lead to a closed symmetry algebra. However, in order for the vector fields to span the corresponding symmetry algebra, their commutator table should be closed. This leads to restrictions on the arbitrary functions of time, and space translation is obtained as one of the spanning vector fields for the associated symmetry algebra, consistent with solitary solutions.

The Euler-Poincar\'e framework and the computations of symmetry groups of partial differential equations is not restricted to dimension one. In particular, in future work, equations of fluid dynamics such as the Euler equations will be addressed in compact manifolds in dimension two and three. Another open challenge that we want to address is how to deal with boundary conditions. The present approach is only valid on manifolds without boundary.

\section*{Acknowledgements}
EL was supported by NWO grant VI.Vidi.213.070.

\bibliographystyle{plainnat}
\bibliography{biblio}

\end{document}